\documentstyle[floats,aps,preprint]{revtex}

\begin{document}
\draft
\preprint{\vbox{ \hbox{SOGANG-HEP 241/98} \hbox{SNUTP 98-089}}}

\title{Manifestly covariant action 
        for symmetric Chern-Simons field theory}

\author{Won Tae Kim\footnote{electronic address:wtkim@ccs.sogang.ac.kr} 
  and Myung Seok Yoon\footnote{electronic address: younms@physics.sogang.ac.kr}}
\address{Department of Physics and Basic Science Research Institute,\\
Sogang University, C.P.O. Box 1142, Seoul 100-611, Korea}
\maketitle
\bigskip
\begin{abstract}
We study a three-dimensional symmetric 
Chern-Simons field theory
with a general covariance and it
turns out that the original Chern-Simons theory is just
a gauge fixed action of the symmetric Chern-Simons theory
whose constraint algebra belongs to fully first class constraint
system. The Abelian Chern-Simons theory with matter coupling
is studied for the construction of anyon operators without
any ordering ambiguity with the help of this symmetric
Chern-Simons action. 
Finally we shall discuss some connections 
between the present symmetric formulation of Chern-Simons theory
and the St\"ukelberg mechanism.        
\end{abstract}
\begin{flushleft}
\end{flushleft}

\newpage
\section{Introduction \hfill{ }}
A Chern-Simons(CS) theory has been enormously studied in the
(2+1)-dimensional quantum field theory \cite{djt} and applied to an
anyon system \cite{bananyon,prs} and the quantum gravity \cite{ct} 
apart from in its own right.
One of the most intriguing feature in the CS theory is due to
the symplectic structure of gauge fields which is related to
a second class constraint system. The second class constraint 
structure in the CS theory does not give a closed constraint algebra 
in Poisson brackets 
even though a local gauge symmetry exists.

To quantize a second class constraint system, the Dirac method
\cite{dir,ht} may be used in the Hamiltonian quantization.
However, Dirac brackets are generically field-dependent, 
nonlocal, and have a serious ordering problem between
field operators. These are problematic and unfavorable in
finding canonically conjugate pairs. 
Once the first class constraint system is realized, the
usual Poisson bracket corresponding to the quantum 
commutator can be used. Quantization in this direction has
been well appreciated in a gauge invariant manner using 
Becci-Rouet-Stora-Tyutin(BRST) symmetry \cite{ht,brst}.
So one might wonder how to convert the second class constraint system 
in the CS theory into a first class one, and what kind of symmetry is
involved in the symplectic structure. 

On the other hand, it would be interesting
to interpret the original CS theory as a gauge fixed version of
a symmetric CS theory similarly to the anomalous gauge
theory in Ref.~\cite{fs}. Then the symplectic structure
of gauge fields naturally appears after unitary gauge fixing.
We expect that this gauge fixing which gives the
symplectic structure in CS theory 
is independent of the local gauge
symmetry. In anomalous gauge theories, in fact, 
it is well known that the algebra of 
first class gauge constraints becomes second class after
quantization.
In the CS theory, however, the origin of second
class constraint is more or less different in that it
is not due to the anomalous breaking of symmetry
but rather the symplectic structure of the CS term. Henceforth the
Gauss' law as a gauge constraint remains as a first class constraint
in the CS theory. 
In this respect, there may exist some differences to convert
the second class constraint system into the first class one.   
This problem has been extensively studied 
in the context of
Batalin-Fradkin-Tyutin(BFT) Hamiltonian
embedding~\cite{bft} in
Refs.~\cite{ban,kp,brr,bcs,kr}. The BFT method also has been applied to
the other interesting physical problems in Ref.~\cite{bnon}.  
However the unitary gauge exists only in the Hamiltonian defined in
the phase space, and the Wess-Zumino(WZ)
type action which makes the first class constraint system depends
on the details of the matter action coupled to CS term.
Since the origin of second class algebra in CS matter theory
is of no relevance to the form of matter coupling, we expect
the WZ type action should be independent of the matter fields in
contrast to the previous result.
So first of all the pure CS theory should be considered, and matter
coupling will be a simple extension.    
Very recently, the second class constraint algebra
of a (2+1)-dimensional Abelian CS term was converted  
into first class one, the 
BFT method has been applied and 
it turns out that 
the symmetry of relevance to the symplectic structure is a
local translational symmetry \cite{kkp}. 
Unfortunately, the final action corresponding to the first class
Hamiltonian system is not generally covariant and how to couple to
matter fields is uncertain. 

In this paper, we study generally covariant (2+1)-dimensional 
symmetric CS theories. The proposed action is manifestly covariant and
the unitary gauge fixing exists in the final action. 
In Sec. 2, we exhibit the constraint structure of the nonAbelian CS theory 
to reveal the second class algebra. 
In Sec. 3, we briefly review some of the recent study on the symmetric
Abelian CS theory for the self-contained manner 
and find out a clue how to convert the second class constraint system
into first class constraint one for the complicated system in the
generally covariant fashion.
We shall use a very simple method in order to convert constraint
algebra through some observations. In Sec. 4, the symmetric 
nonAbelian CS theory whose constraint algebra is 
fully first class is presented and the usual Poisson brackets are well
defined without recourse to Dirac brackets.
The proposed action is generally covariant and has an additional
local translational symmetry which is of relevance to the
symplectic structure of the CS term. 
In Sec. 5, as an field-theoretic application, anyon operators are 
constructed by using the symmetric CS term coupled to complex 
scalar fields. After all, we find the anyon operator without
resorting to the Dirac brackets and naturally circumvent the 
ordering ambiguity between gauge fields. 
Finally we discuss the present method and its connection to
the St\"ukelberg mechanism which is apparently unrelated to our analysis
in Sec. 6. 

\section{Constraint structure of 2+1 dimensional CS theory \hfill{ }}

\setcounter{equation}{0}
We now exhibit some of the salient features of
the 2+1 dimensional nonAbelian CS theory whose
action is given by
\begin{equation} \label{S:cs}
  S_{\rm CS} = \kappa \int d^3 x \/ \epsilon^{\mu \nu \rho}
  {\rm tr}(A_{\mu}\partial_{\nu}A_{\rho} - \frac{2}{3}{\it i}
  A_{\mu}A_{\nu}A_{\rho}),
\end{equation}
where the diagonal metric $g_{\mu \nu}=diag(+,-,-)$ and
$\epsilon^{012}=+1$. The Lie algebra-valued gauge field is defined by 
$A_{\mu}=A^a_{\mu}T^a$  satisfying
$[T^a, T^b]={\it i}f^{abc}T^c$ and ${\rm tr}( T^a
T^b)=\frac{1}{2}\delta^{ab}$ where $T^a$ is a Hermitian generator.
Note that the large gauge invariance of the nonAbelian Chern-Simons
action  requires
the quantization of the dimensionless constant $\kappa$,~$\kappa=
\frac{n}{4\pi}(n \in Z)$.
The canonical momenta of the action~(\ref{S:cs}) are given by 
\begin{eqnarray}
  & &\Pi^{0a} \approx 0, \label{motime}\\
  & &\Pi^{ia} = \frac{\kappa}{2}\epsilon^{ij}A_j^a, \label{Cmom}
\end{eqnarray}
which are all primary constraints. Performing the Legendre transformation,
the canonical Hamiltonian is written as 
\begin{equation}\label{Cham}
  H_c = - \int d^2 {\bf x} \/ A^{0a}(\partial_i \Pi^{ia} +
  \frac{\kappa}{2}\epsilon^{ij}\partial_iA_j^a +
  \frac{\kappa}{2} f^{abc} \epsilon^{ij} A^b_iA^c_j).
\end{equation}
At this stage, we define nonvanishing Poisson brackets as
\begin{equation}
\{A^a_\mu (x), \Pi^{\nu b} (y) \}= g_\mu^\nu \delta^{ab}
\delta^2 ({\bf x} - {\bf y}).
\end{equation}
The time evolution of the primary constraint Eq.~(\ref{motime})
yields Gauss' law constraint as a secondary constraint, which is simply
written by
\begin{equation}\label{Gauss}
  \partial_i \Pi^{ia} +
  \frac{\kappa}{2}\epsilon^{ij}\partial_iA_j^a +
  \frac{\kappa}{2} f^{abc} \epsilon^{ij} A^b_iA^c_j \approx 0.
\end{equation}
Further time evolution of the Gauss' constraint (\ref{Gauss}) gives
no more additional constraint.
We therefore obtain the following set of constraints
\begin{eqnarray}
  \Omega^{0a} &=& \Pi^{0a} \approx 0, \label{Cmom1}\\
  \Omega^{ia} &=& \Pi^{ia} - \frac{\kappa}{2}\epsilon^{ij}A_j^a
     \approx 0, \label{Cmom2} \\
  \Omega^a &=& \partial_i \Pi^{ia} + f^{abc} A^b_i \Pi^{ic}
     + \frac{\kappa}{2}\epsilon^{ij}\partial_iA^a_j \approx 0.
     \label{Cmom3} 
\end{eqnarray}
By using the Poisson bracket, the first class 
constraint algebra is given by
\begin{eqnarray}
  & &\{\Omega^{0a}(x),\Omega^{0b}(y)\} =
    \{\Omega^{0a}(x),\Omega^b(y)\} \approx 0, \nonumber\\
  & &\{\Omega^a(x), \Omega^b(y)\} = f^{abc} \Omega^c(x)
    \delta^2 ({\bf x} - {\bf y}), \label{PB1}
\end{eqnarray}
while the nonvanishing second class algebra is written as 
\begin{equation}
  \{\Omega^{ia}(x),\Omega^{jb}(y)\} = -\kappa\epsilon^{ij} \delta^{ab}
  \delta^2 ({\bf x} - {\bf y}). \label{PB2}
\end{equation}
Note that the second class constraint algebra~(\ref{PB2}) 
is reminiscent of an anomalous commutator of the anomalous 
gauge theory which reflects a local gauge symmetry breaking,
while the first class constraints (\ref{Cmom1})
and (\ref{Cmom3}) guarantee the local gauge symmetry.
So one might wonder what kind of additional local symmetry is
broken in the second class constraint algebra~(\ref{PB2}).
Therefore, it is necessary to study how to convert the second class
constraint algebra into the first class one to answer this
question. 

There exist largely two ways to recover the symmetry in
an extended configuration space.
One is the St\"ukelberg mechanism \cite{stu} which is done by
performing a gauge transformation, $A_\mu \rightarrow
U^{-1} A_\mu U - i U^{-1} \partial_\mu U$ and identifying
a new field $U$ as a St\"ukelberg scalar field.
The other one is the BFT method which converts a
second class constraint algebra into a first 
class constraint algebra in the Hamiltonian formalism
by introducing new conjugate pairs.
Unfortunately, in the nonAbelian CS theory these
methods may not be successful so far since we do not know
what is the relevant symmetry to this kind of constraint
algebra (\ref{PB2}) for the case of the St\"ukelberg mechanism, 
and there are
some arbitrariness to introduce conjugate fields
for the BFT formalism. So the general covariance
of the action is lost in the course of calculation in
the latter formalism~\cite{kkp}. 
Therefore we suggest a method to convert the
second class constraint system into a first class
one in a generally covariant fashion by
simply substituting the original gauge field by a new field.
This method as a matter of fact amounts to the St\"ukelberg mechanism
which will be discussed in later. 
In Sec. 4 and 5, we shall consider this
method, and apply to the (2+1)-dimensional nonAbelian
CS theories and CS matter coupling.

\section{2+1 dimensional symmetric Abelian CS theory \hfill{ }}
\setcounter{equation}{0}
In an Abelian CS theory, some new results are obtained in
recent work \cite{kkp} on the symmetry of relevance to
the symplectic structure of the CS theory. At this juncture we
recapitulate some of the results and will assume a symmetry
in order to apply nonAbelian cases 
in a generally covariant fashion.

One can now apply a BFT Hamiltonian embedding of the CS term   
by introducing auxiliary fields \cite{bft}.
The starting (2+1)-dimensional Abelian CS 
Lagrangian is given by 
\begin{equation}
\label{13}
{\cal L}_{(0)}  =
             \frac{\kappa}{2}\epsilon^{\mu\nu\rho}A_\mu^{(0)}
                     \partial_\nu A_\rho^{(0)},
\end{equation}
where $A_\mu ^{(0)}$ is an original CS gauge field and for simplicity
the CS coefficient is set to $\kappa=1$.
We introduce an auxiliary field $A_i^{(1)}$ satisfying
\begin{equation}
\label{problem}
\{A_i^{(1)}(x),A_j^{(1)}(y)\}=\vartheta_{ij}(x,y) 
\end{equation}
which makes the second
class constraint $\omega^{i} = \Pi^{i} - \frac{\kappa}{2}
\epsilon^{ij} A_j \approx 0$ into a first class one where
$\vartheta_{ij}(x,y)$ will be explicitly determined in later.
Making use of the auxiliary field $A_i^{(1)}$, we could write the effective
first class constraints as $\tilde{\omega}^i
(\pi_{(0)}^\mu,A_\mu^{(0)};A_i^{(1)})  = \omega^i +
\sum_{n=1}^\infty C_{(n)}^i$ satisfying the boundary condition 
$\tilde{\omega}^i(\pi_{(0)}^\mu,A_\mu^{(0)};0)=\omega^i$ 
as well as requiring the strong involution, {\it i.e.,} 
$\{\tilde{\omega}^i,\tilde{\omega}^j\}=0$.
Here $C_{(n)}^i$ is assumed to be proportional to $(A_i^{(1)})^n$.
In particular, the first order correction in these infinite series is given by
\begin{equation}
\label{16}
C_{(1)}^i(x) = \int \,d^2{\bf y}\/ X^{ij}(x,y)A_j^{(1)}(y),
\end{equation}
and the requirement of the strong involution gives the condition 
\begin{equation}
\label{17}
-\kappa \epsilon_{ij}\delta^2 ({\bf x}-{\bf y})+
\int \, d^2{\bf u}d^2{\bf v} \/
X^{ik}(x,u)\vartheta_{k\ell}(u,v)X^{j\ell}(v,y)=0.
\end{equation}
We take the simple solution of $\vartheta_{ij}$ and $X^{ij}$ as
\begin{eqnarray}
\label{theta}
\vartheta_{ij}(x,y) &=& \epsilon_{ij}\delta^2 ({\bf x}-{\bf y}), \\
\label{X}
X^{ij}(x,y) &=& -\epsilon^{ij}\delta^2({\bf x} - {\bf y}).
\end{eqnarray}
By using Eqs. (\ref{theta}) and (\ref{X}),
we obtain the strongly involutive first class constraints 
which are proportional only to the first order of the auxiliary field as
\begin{equation}\label{20}
  \tilde{\omega}_{(0)}^i = \pi_{(0)}^i - \frac12 \epsilon^{ij}A_j^{(0)}
  - \epsilon^{ij}A_j^{(1)}=0,
\end{equation}
and the canonical Hamiltonian density
\begin{equation}
\label{H_c}
\tilde{\cal H}_c =  - A_0^{(0)}\epsilon^{ij}\partial_i
                \left(A_j^{(0)} + A_j^{(1)}\right),
\end{equation}
satisfying $\{\tilde{\omega}^i, \tilde{H}_c\}=0$.
The corresponding Lagrangian to Eq. (\ref{H_c}) with the auxiliary
field $A_i^{(1)}$ is obtained through the usual Legendre
transformation \cite{ban,kp,brr,bcs,kr} as
\begin{eqnarray}
\label{L_1}
{\cal L}_{(1)}&=&-\frac12\epsilon^{ij} A_i^{(0)}\dot{A}_j^{(0)}
                 +  A_0^{(0)}\epsilon^{ij}\partial_iA_j^{(0)} \nonumber\\
              && - \frac{1}{2}\epsilon^{ij} A_i^{(1)}\dot{A}_j^{(1)}
                  + A_0^{(0)} \epsilon^{ij} \partial_i A_j^{(1)} 
                  -\epsilon^{ij}A_i^{(1)}\dot{A}_j^{(0)}.
\end{eqnarray}
However, the first iteration of the BFT formalism is not satisfactory 
since the action (\ref{L_1}) is not the genuine first class constraint
system in the Poisson algebra, 
which is seen from the following reconsideration of 
Hamiltonian
analysis. 
The canonical momenta from (\ref{L_1}) are $\pi_{(0)}^0=0$, 
$\pi_{(0)}^i=\frac12\epsilon^{ij}A_j^{(0)} + \epsilon^{ij}A_j^{(1)}$, and
$\pi_{(1)}^i=\frac{1}{2}\epsilon^{ij}A_j^{(1)}$. 
>From the time stability conditions of these primary constraints, 
we can get one more secondary constraint and after redefining
the constraints we can easily obtain the maximally irreducible first class 
constraints as
$\omega^0=\pi_{(0)}^0 \approx 0$,
$\omega^3 = \partial_i\pi_{(0)}^i 
              + \frac12\epsilon^{ij}\partial_iA_j^{(0)} \approx 0$,
and
\begin{eqnarray}\label{24}
\tilde\omega_{(1)}^i=\pi_{(0)}^i - \frac{1}{2}\epsilon^{ij}A_j^{(0)}      
- (\pi_{(1)}^i+\frac{1}{2}\epsilon^{ij}A_j^{(1)}) \approx 0,
\end{eqnarray}
as well as the problematic constraint 
\begin{eqnarray}\label{da}
\omega_{(1)}^i = \pi_{(1)}^i-\frac{1}{2}\epsilon^{ij}A_j^{(1)} \approx 0,
\end{eqnarray}
which is unfortunately second class.
Therefore there remains still a second class constraint even
after the first order of correction. 
The consistent bracket is defined by the Dirac bracket as
\begin{equation}
\label{26}
\{A_i^{(1)}(x), A_j^{(1)}(y)\}_D = \epsilon_{ij}\delta^2({\bf x} - {\bf y}),
\end{equation}
which is compatible with Eqs.(\ref{problem}) and (\ref{theta}) 
to make the second class constraint $\omega^i$ into the first class one. 
Therefore the bracket defined in Eq. (\ref{problem}) is not 
the Poisson bracket but the Dirac one. This is reason why
we do not obtain the first class constraint system in the Poisson
algebra. 
This enforces the BFT Hamiltonian embedding of the CS theory not stopping any 
finite number of steps. 
Therefore the same step should be infinitely repeated until  
fully first class constraint algebra appears 
by introducing infinite auxiliary fields denoted by $A_i^{(n)}$.
The similar circumstances are already encountered in 
self-dual theory as a chiral boson theory, 
\cite{wot,mwy}.
In this respect, all the previous results \cite{ban,kp,brr,bcs,kr}
of the BFT formalism applied to the CS matter coupling  cases 
are also confronted with this kind of problem.
After repeating the BFT formalism infinitely, the final 
action can be written in the form 
\begin{eqnarray}
  {\cal L}_{\rm SCS} &=& -\frac{1}{2} \epsilon^{ij} A_i^{(0)} \dot{A}_j^{(0)}
      + A_0^{(0)} \epsilon^{ij} \partial_i A_j^{(0)} - \frac{1}{2}
      \epsilon^{ij} \sum^{\infty}_{n=1} A_i^{(n)} \dot{A}_j^{(n)} +
      A_0^{(0)} \epsilon^{ij} \sum^{\infty}_{n=1} \partial_i A_j^{(n)}
      \nonumber\\ 
  & & - \epsilon^{ij} \sum^{\infty}_{n=1} A_i^{(n)} \dot A_j^{(0)}
      - \epsilon^{ij} \sum^{\infty}_{n=1} \sum^{\infty}_{m=n+1}
      A_i^{(m)} \dot{A}_j^{(n)}. \label{new}
\end{eqnarray}
Then the symmetric CS theory (\ref{new})
with the infinite number of auxiliary fields now completely 
gives the first class constraint system 
and strongly vanishing Poisson brackets between 
constraints in contrast to the finite iteration of
BFT method. 
Remarkably the other convenient action of Eq. (\ref{new}) is written in
the compact form as
\begin{equation}\label{newcom}
  {\cal L}_{\rm SCS} = - \frac{1}{2} \epsilon^{ij} \left( A_i^{(0)} +
    \sum^{\infty}_{n=1} A_i^{(n)} \right) \left( \dot{A}_j^{(0)} +
    \sum^{\infty}_{n=1} \dot{A}_j^{(n)} \right) + A_0^{(0)}
    \epsilon^{ij} \partial_i \left( A_j^{(0)} +
    \sum^{\infty}_{n=1} A_j^{(n)} \right)
\end{equation}
after some resummations of auxiliary fields.
This action is invariant under the following local gauge 
transformation
\begin{eqnarray}
\label{tr}
\delta A_0^{(0)}&=&\partial_0\Lambda,\nonumber\\
\delta A_i^{(0)}&=&\partial_i\Lambda
            +\epsilon_i^{(1)}, \nonumber\\
\delta A_i^{(n)}&=&-\epsilon_i^{(n)}+\epsilon_i^{(n+1)} (n=1,2, \dots).
\end{eqnarray}
The transformation rule (\ref{tr}) is implemented by
the usual U(1) gauge transformation with the gauge parameter $\Lambda$
and a new type of local symmetry with $\epsilon_i^{(n)}$. 
Note that the procedure to arrive the final result is
cumbersome if one want to apply it to other case as an nonAbelian
extension of the CS term. After all, we have
obtained the first class constraint system from the second 
class original CS theory. Further the general covariance is lost in
the course of our calculation in BFT Hamiltonian embedding.
In fact, the general covariance of CS theory is an essential
feature of the CS theory, which is a metric independent property.
Therefore we overcome these problems with the help of
some observations in the next section.

\section{2+1 dimensional symmetric nonAbelian CS theory \hfill{ }}
\setcounter{equation}{0}
 In this section, we shall generalize the previous Abelian result 
to the nonAbelian CS theory with maintaining general
covariance. We observe that the time component of gauge fields
is missing in the transformation rule (\ref{tr}), which
becomes in fact a fundamental reason why we did not obtain the
generally covariant first class constraint system.
The resolution of the covariance problem
does not appear in a natural way in 
the BFT Hamiltonian embedding of our model.
Furthermore nonAbelian extension of the previous Abelian
result within the BFT formalism may be intractable 
and seems to be cumbersome because of some complexities.
Therefore, without further resort to the BFT formalism, 
at this stage we simply assume that the
new local translational symmetry combined with the local
Abelian gauge symmetry is promoted to the following form 
\begin{eqnarray}
  \delta A_\mu^{(0)} &=& \tilde{D}_\mu \epsilon^{(0)} + \epsilon_\mu^{(1)},
  \label{delta-0} \\
  \delta A_\mu^{(n)} &=& -\epsilon_\mu^{(n)} + \epsilon_\mu^{(n+1)},
  \qquad (n=1,2,\cdots)
   \label{delta-n}
\end{eqnarray}
where the covariant derivative is defined by 
$\tilde{D}_\mu \epsilon^{(0)} = \partial_\mu
\epsilon^{(0)} - i [\tilde{A}_\mu^{(n)},\epsilon^{(0)}]$ and
$\tilde{A}_\mu = A_\mu^{(0)} + \sum_{n=1}^\infty A_\mu^{(n)}$.
The matrix valued gauge and translational parameters are denoted by
$\epsilon^{(0)}$ and $\epsilon_\mu^{(n)} $ respectively. 
Note that the covariant derivative is expressed in terms 
of not only an original CS gauge field but also auxiliary fields.
Then the symmetric action under the transformation (\ref{delta-0}) 
and (\ref{delta-n}) is given by
\begin{equation}\label{L:2+1.cs}
  {\cal L}_{\rm SCS} = \kappa \, \epsilon^{\mu\nu\rho}\, {\rm tr} \left(
    \tilde{A}_\mu \partial_\nu \tilde{A}_\rho - \frac23 i
    \tilde{A}_\mu \tilde{A}_\nu \tilde{A}_\rho \right).
\end{equation}
It is interesting to note that the symmetric CS action is
apparently the same form as the original CS action except for
a new gauge field which is composed of the infinite number of  vector
fields. Under the given local transformation (\ref{delta-0}) and 
(\ref{delta-n}), the Lagrangian~(\ref{L:2+1.cs}) is invariant up to a
total divergence term as
\begin{equation}\label{transf:L}
  \delta {\cal L}_{\rm SCS} = \partial_\mu \left[ \kappa\epsilon^{\mu\nu\rho}\,
  {\rm tr}\/ \left( \tilde{A}_\nu \tilde{D}_\rho \epsilon^{(0)} - 2i
  \tilde{A}_\nu \tilde{A}_\rho \epsilon^{(0)} \right) \right].
\end{equation}
Note that the above total divergence does not depend on the
translational parameter and the sufficient convergence condition
of gauge group parameter $\epsilon^{(0)}$ guarantees the
invariance of the action. The two symmetries are controlled
by the independent parameters and the symmetry transformation
rules can be arbitrarily separated by a modified gauge transformation 
\begin{equation}\label{gauge}
\delta A_\mu^{(0)} = \tilde{D}_\mu \epsilon^{(0)}, \qquad 
\delta A_\mu^{(n)}= 0
\end{equation}
and a translational symmetry 
\begin{equation}\label{tans}
\delta A_\mu^{(0)} =  \epsilon_\mu^{(1)},\qquad \delta A_\mu^{(n)} 
     = -\epsilon_\mu^{(n)} + \epsilon_\mu^{(n+1)} \qquad (n=1,2,\cdots)
\end{equation}
respectively. We should recall that the
infinite number of vector fields are involved in the
covariant derivative which is unusual.

On the other hand, the collective expression of the transformations
from Eqs. (\ref{gauge}) and (\ref{tans}) is written as   
\begin{equation}\label{transf:til}
\delta \tilde{A}_\mu = \tilde{D}_\mu \epsilon^{(0)}
\end{equation}
and the symmetric action (\ref{L:2+1.cs}) is automatically
invariant under the transformation rule. This situation is
very similar to the usual gauge invariance of the original
CS term. Note that the concise expression (\ref{transf:til})
may not be decomposed into the transformations (\ref{delta-0}) 
and (\ref{delta-n}) since the decomposition is not unique. Hence the collective
expression (\ref{transf:til}) is just only for convenience.  
As for the finite transformation of $\tilde{A}_\mu $,
it is naturally written as $\tilde{A}_\mu \rightarrow
U^{-1} \tilde{A}_\mu U -i U^{-1} \partial_\mu U$ where
the finite transformation matrix is $U=e^{iT^a \epsilon^{(0)a}}$
and the quantization condition of CS coefficient is still valid.
In our consideration, the general covariance has been
maintained.

To check whether or not the symmetric CS action (\ref{L:2+1.cs}) gives
a desired first class constraint system, the canonical momenta are
derived from Eq.~(\ref{L:2+1.cs}) 
\begin{eqnarray}
  \Pi_{(n)}^0 &\approx& 0, \label{mom:0} \\
  \Pi_{(n)}^i &=& \frac{\kappa}{2} \epsilon^{ij} \tilde{A}_j,
  \label{mom:i}  
\end{eqnarray}
where the spatial momentum is a collection of all vector fields.
So the primary Hamiltonian becomes
\begin{equation}
  H_p = \int \, d^2 {\bf x} \/ \Big[ - \tilde{A}_0^a \left( 
    \kappa\epsilon^{ij} \partial_i \tilde{A}_j^a + \frac12 f^{abc}
    \tilde{A}_i^b \tilde{A}_j^c \right) + \sum_{n=0}^\infty
    \lambda_\mu^{(n)a} \Omega_{(n)}^{\mu a} \Big],
\end{equation}
where the two constraints are rewritten as for convenience
\begin{eqnarray}
  \Omega_{(n)}^{0a} &=& \Pi_{(n)}^{0a} \approx 0 \label{const:0} \\
  \Omega_{(n)}^{ia} &=& \Pi_{(n)}^{ia} - \Pi_{(n+1)}^{ia} \approx 0,
  \label{const:i} 
\end{eqnarray}
where hereafter we assume $n=0,1,2,\cdots$.
The Gauss' law is given by
\begin{equation}\label{const:gauss}
  \Omega^{3a} = ( \tilde{D}_i \Pi_{(0)}^i )^a + \frac{\kappa}{2}
  \epsilon^{ij}\partial_i \tilde{A}_j^a \approx 0.
\end{equation} 
At first sight, the primary constraint (\ref{mom:i}) and the Gauss
constraint (\ref{const:gauss}) seem to be a second class, however it
is actually first class one as easily seen from the redefined form of
Eq. (\ref{const:i}).

To make this explicit in another way, we now rewrite the action
(\ref{L:2+1.cs}) after some resummations of terms in the action, which
is given by
\begin{eqnarray}
& &  {\cal L}_{\rm SCS}=\sum_{n=0}^{\infty}
\left( \frac{\kappa}{2}\epsilon^{ij}A_j^{(n)a} +
  \kappa\epsilon^{ij}\sum_{m=n+1}^{\infty}A_j^{(m)a} \right)
\dot{A}_i^{(n)a} +
\sum_{n=0}^{\infty}A_0^{(n)a}\left[\sum_{m=0}^{\infty} \kappa
  \epsilon^{ij}\partial_iA_j^{(m)a} \right. \nonumber \\
& & \left.+\sum_{m=0}^{\infty} f^{abc} 
\left( \frac{\kappa}{2}\epsilon^{ij}A_j^{(m)c} +
  \kappa\epsilon^{ij}\sum_{l=m+1}^{\infty}A_j^{(l)c}\right)A_i^{(m)b}\right].
\label{L:total1} 
\end{eqnarray}
Then the canonical momenta for $A_\mu^{(n)a}$ are
\begin{eqnarray}
  \Pi_{(n)}^{0a} &\approx& 0, \label{mom-0} \\
  \Pi_{(n)}^{ia} &=& \frac{\kappa}{2} \epsilon^{ij} A_j^{(n)a} +
  \kappa \epsilon^{ij} \sum_{m=n+1}^\infty A_j^{(m)a}, \label{mom-i}
\end{eqnarray}
and the primary Hamiltonian is  
\begin{equation}\label{H}
  H_p = \int d^2 {\bf x} \, \left[ -\sum_{n=0}^\infty A_0^{(n)a}
  \left( \partial_i \Pi_{(0)}^{ia} + f^{abc} \sum_{m=0}^\infty
  A_i^{(m)b} \Pi_{(m)}^{ic} + \frac{\kappa}{2} \epsilon^{ij}
  \partial_i A_j^{(0)a} \right) + \sum_{n=0}^\infty \lambda_\mu^{(n)a}
  \Omega_{(n)}^{\mu a} \right].
\end{equation}
The constraints are written as
\begin{eqnarray}
  \Omega_{(n)}^{0a} &=& \Pi_{(n)}^{0a} \approx 0, \label{const-0} \\
 \Omega_{(n)}^{ia} &=& \Pi_{(n)}^{ia} - \frac{\kappa}{2}
      \epsilon^{ij} A_j^{(n)a} - \left( \Pi_{(n+1)}^{ia} + \frac{\kappa}{2}
      \epsilon^{ij} A_j^{(n+1)a} \right) \approx 0, \label{const-i} \\
  \Omega^{3a} &=& \partial_i \Pi_{(0)}^{ia} + f^{abc}
      \sum_{n=0}^\infty A_i^{(n)b} \Pi_{(n)}^{ic} + \frac{\kappa}{2}
      \epsilon^{ij} \partial_i A_j^{(0)a} \approx 0. \label{const-3}
\end{eqnarray}
Note that the constraint (\ref{const-i}) is obtained from a
recombination and the last term in Eq. (\ref{mom-i}) is written
by the next order of Eq. (\ref{mom-i}). The Poisson brackets between the
constraints yield the desired first class constraint algebra after
some calculations,
\begin{eqnarray}
  & & \{ \Omega_{(n)}^{0a} (x),\Omega_{(m)}^{0b} (y) \} = 
      \{ \Omega_{(n)}^{0a} (x), \Omega_{(m)}^{ib} (y) \} =
      \{ \Omega_{(n)}^{0a} (x), \Omega^{3b} (y) \} \approx 0, \label{PB-1} \\
  & & \{ \Omega_{(n)}^{ia} (x), \Omega_{(m)}^{jb} (y) \} \approx 0,
      \label{PB-2} \\
  & & \{ \Omega_{(n)}^{ia} (x), \Omega^{3b} (y) \} = f^{abc}
      \Omega_{(n)}^{ic} (x) \delta^2 ({\bf x} - {\bf y}), \label{PB-3} \\
  & & \{ \Omega^{3a} (x), \Omega^{3b} (y) \} = f^{abc} \Omega^{3c} (x)
      \delta^2 ({\bf x} - {\bf y}), \label{PB-4}
\end{eqnarray}
where we used the Jacobi identity to show
the last algebra Eq.(\ref{PB-4}). 

As a result, the symmetric nonAbelian CS theory is obtained
by simply redefining the original gauge field as the
new tilde field. The constraint algebra is fully first
class. If one chooses an unitary gauge for the translational symmetry
$A_\mu^{(1)} = A_\mu^{(2)} = \cdots = 0$,
then the original CS theory recovers and the consistent bracket will
be the Dirac bracket as (\ref{PB2}). Further gauge fixing corresponding
to the usual Coulomb type gauge fixing, the constraint algebra becomes
fully second class.   

\section{Construction of anyon operators \hfill{ }}
\setcounter{equation}{0}
The Abelian Chern-Simons theory coupled to the complex
matter fields is reconsidered in our formalism, which is
essentially first class constraint constraint system.  
By analyzing this model without 
any gauge fixing condition, 
we naturally obtain gauge-independent anyon operators
which is also free from ordering problems between field operators.

The matter coupling to the CS term is given by the action 
\begin{equation}
\label{original}
{\cal L}_0 = {\cal L}_{\rm CS}
( A_\mu^{(0)} ) + (\partial_\mu\phi + 
i A_\mu^{(0)} \phi)^\dagger (\partial_\mu\phi + i A_\mu^{(0)} \phi ),
\end{equation}
where it is a second class constraint system 
already studied in Ref. \cite{ban,kp} in terms of 
the BFT Hamiltonian
embedding. In these works, the first class system was
impossible when we assume the usual Poisson brackets.
If one wants to quantize the system by using the
Poisson bracket(commutator) instead of Dirac ones,
then the symmetric action will be adopted by substituting
$A_\mu^{(0)}$ by $\tilde{A}_\mu$ in the given action, which
is simply written as
\begin{eqnarray*}
{\cal L} = {\cal L}_{\rm SCS} (\tilde{A}_\mu) 
+ (\tilde{D}^\mu\phi)^*(\tilde{D}_\mu\phi),
\end{eqnarray*}
where the covariant derivative is defined by $\tilde{D}_\mu \phi =
\partial_\mu \phi + ie \sum_{n=0}^\infty A_\mu^{(n)} \phi$.
This original action (\ref{original})
can be recovered at the Lagrangian stage by choosing unitary gauge
$A^{(1)}_\mu=A^{(2)}_\mu= \cdots =0$. If we turns off the matter
fields, then the pure symmetric CS theory appears. So our formulation
on CS field theory is independent of matter contents, which is in
contrast with the previous result.

The Lagrangian with the Klein-Gordon fields $\phi$ and $\phi^*$ is given by
\begin{equation}\label{csm}
  {\cal L} = \frac{\kappa}{2} \epsilon^{\mu\nu\rho} \sum_{n=0}^\infty
    \left( \sum_{m=0}^n A_\mu^{(n)} \partial_\nu A_\rho^{(m)} +
    \sum_{m=n+1}^\infty A_\mu^{(m)} 
    \partial_\nu A_\rho^{(n)} \right)
    + (\tilde{D}^\mu\phi)^*(\tilde{D}_\mu\phi),
\end{equation}
through the resummations as Eq.~(\ref{L:total1}). 
>From this Lagrangian (\ref{csm}), 
the canonical momenta
are obtained as 
\begin{eqnarray}
  & & \Pi_{(n)}^0 \approx 0, \label{mom:0}\\
  & & \Pi_{(n)}^i = \frac{\kappa}{2} \epsilon^{ij}A_j^{(n)} + \kappa
      \epsilon^{ij} \sum_{m=n+1}^\infty A_j^{(m)}, \label{mom:i}\\
  & & \Pi = (\tilde{D}_0 \phi)^* , \\
  & & \Pi^* = (\tilde{D}_0 \phi),
\end{eqnarray}
where $\Pi_{(n)}^0$, $\Pi_{(n)}^i$, $\Pi$, and $\Pi^*$ are
the conjugate momenta of $A_0^{(n)}$, $A_i^{(n)}$, $\phi$, and
$\phi^*$, respectively. From the Legendre transformation,
we obtain the canonical Hamiltonian
\begin{equation}\label{eq:H_c}
  {\cal H}_{\rm c} = |\Pi|^2 + |{\bf D}\phi|^2  -
  \sum_{n=0}^\infty A_0^{(n)} G,
\end{equation}
and Gauss' law is written as
\begin{equation}
  \label{eq:G}
  G = \kappa\epsilon^{ij} \sum_{n=0}^\infty \partial_i A_j^{(n)} + J_0,
\end{equation}
where the source current  
\begin{equation}\label{eq:J}
  J_\mu = ie \left[ (\tilde{D}_\mu \phi)^* \phi - (\tilde{D}_\mu \phi)
  \phi^* \right]
\end{equation}
is conserved. The primary constraints are
\begin{eqnarray}
  \Omega_{(n)}^0 &=& \Pi_{(n)}^0 \approx 0 \label{const:0}\\
  \Omega_{(n)}^i &=& \Pi_{(n)}^i - \frac{\kappa}{2}
      \epsilon^{ij} A_j^{(n)} - \left( \Pi_{(n+1)}^i + \frac{\kappa}{2}
      \epsilon^{ij} A_j^{(n+1)} \right) \approx 0, \label{const:i} 
\end{eqnarray}
where the constraints (\ref{const:i}) are obtained by the
recombination of the momenta (\ref{mom:i}). From the Gauss' law
(\ref{eq:G}) as a secondary constraint is rewritten as
\begin{equation}\label{const:3} 
  \Omega^3 = \partial_i \Pi_{(0)}^i + \frac{\kappa}{2}
  \epsilon^{ij} \partial_i A_j^{(0)} + ie\left( \Pi \phi -
  \Pi^* \phi^*\right) \approx 0. 
\end{equation}
The fundamental Poisson brackets are defined by
\begin{eqnarray}
  & & \{A_\mu^{(n)} (x), \Pi_{(m)}^\nu (y) \} =\delta_{nm} 
      g_\mu^\nu \delta^2 ({\bf x} - {\bf y}), \nonumber\\
  & & \{\phi(x), \Pi(y) \} = \{ \phi^*(x), \Pi^*(y) \} 
      = \delta^2 ({\bf x} - {\bf y}). \label{PB}
\end{eqnarray}
Note that all the constraints $\Omega^{0},~~\Omega^{i}$, and
$\Omega^3$ are first class and the problematic
algebra (\ref{problem}) does not appear in our calculation,
therefore the usual Poisson brackets can be used 
in the construction of anyon operators.

Following the procedure suggested in Ref. \cite{bananyon},
a gauge-invariant anyon operator denoted by $\hat{\phi}$ is 
constructed as
\begin{equation}
  \label{anyon}
  \hat{\phi} (x) = \exp \left( \int\/dy \,2is(\theta)\omega(x-y) J_0(y) +
  i\int_{x_0}^x \/ dy_i \, {\tilde A}_i (y) \right) \phi(x),
\end{equation}
satisfying 
\begin{equation}\label{eq:newphi}
  [J_0(x), \hat{\phi}(y)] = \hat{\phi}(y) \delta^2 ({\bf x} - {\bf y}),
\end{equation}
where $\Omega$ is given by
\begin{equation}
  \omega(x-y) = \arctan \left( \frac{x^2 - y^2}{x^1 - y^1} \right)
\end{equation}
which is multivalued and statistics is characterized by
$s (\theta) = 1/\theta$ for the
Klein-Gordon field.
Therefore, we can show that
\begin{equation}
  \label{eq:hat}
  \hat{\phi}(x) \hat{\phi}(y) = e^{\pm 2i s (\theta)} \hat{\phi}(y)
  \hat{\phi}(x),
\end{equation}
using $\omega(x-y) - \omega(y-x) = \pm \pi$.
Note that we have not assumed any path-ordering of gauge field in
Eq. (\ref{anyon}) since the gauge fields themselves are commuting
in contrast to the conventional derivation of anyon operators.
We have shown that the anyon operators are constructed in terms of
the Poisson brackets in the enlarged field space. On the other
hand, the Hamiltonian formalism and anyon operators are studied
in many literatures so far \cite{prs} in the reduced field space,
and it seems to be regarded the anyon system as an effective theory
when the phase space is reduced. In our formulation, the construction
of anyon operator is possible in the gauge-independent way and 
without any ordering problems.  
\section{Discussion \hfill{ }}

Now it seems to be appropriate to comment on our symmetric 
action in the context of the St\"ukelberg mechanism.
Our derivation of the symmetric CS theory relies on a conjecture from
the Abelian BFT method in some sense. One might wonder how to derive
the symmetric CS action in terms of St\"ukelberg mechanism. 
Here we briefly discuss how to obtain the symmetric 
action in this method. The internal gauge parameter of no relevance to
our discussion is definitely independent of the translational symmetry
parameter and we can simply write the transformation as $A_\mu^{(0)}
\rightarrow A_\mu^{(0)}+ \epsilon_\mu^{(1)}$ with simply setting
$\epsilon^{(0)} = 0$. The above transformation rule is
derived as usual directly from the spatial integration of constraint
Eqs.~(\ref{Cmom1}) and (\ref{Cmom2}) which is given by
a symmetry generator expressed as
$Q^{(0)} = -2 \int \/ d^2 {\bf x}\, {\rm tr} \/ [\epsilon_0^{(1)}
\Pi_{(0)}^0 + \epsilon_i^{(1)} (\Pi_{(0)}^i - \frac12 \epsilon^{ij}
A_j^{(0)})]$.
The Poisson bracket of gauge fields with this generator $Q^{(0)}$ yields
the above transformation rule, however the original CS action is not
invariant under this transformation. 
To make the CS action invariant under this transformation rule
according to the St\"ukelberg mechanism, we substitute the original field 
$A_\mu^{(0)}$ with $A_\mu^{(0)}+ \epsilon_\mu^{(1)}$ and identify
$\epsilon_\mu^{(1)}$ as a new vector field $A^{(1)}_\mu$. 
Unfortunately, the transformed CS action which partially corresponds
to the first iterated action (\ref{L_1}) of BFT Hamiltonian embedding
does not give the first class constraint system, which is easily
checked with the help of constraint analysis. 
Therefore the second step of St\"ukelberg mechanism similar to 
the above one is needed. In this second step, from the transformed
action one can obtain the constraints as $\Pi_{(1)}^{0a}=
\Pi_{(1)}^{1a} \approx 0 $,
$\Pi_{(0)}^{ia} - \frac{\kappa}{2} \epsilon^{ij} A_j^{(0)a} -
(\Pi_{(1)}^{ia} + \frac{\kappa}{2} \epsilon^{ij} A_j^{(1)a} ) \approx
0$, and $\Pi_{(1)}^{ia} - \frac{\kappa}{2} \epsilon^{ij}
A_j^{(1)a}\approx 0$ 
which gives the following transformation rule as $A_\mu^{(0)}
\rightarrow A_\mu^{(0)}+ \epsilon_\mu^{(1)}$ and $A_\mu^{(1)}
\rightarrow A_\mu^{(1)} - \epsilon_\mu^{(1)}+\epsilon_\mu^{(2)}$ in 
terms of the generator 
$Q^{(1)} = -2 \int \/ d^2 {\bf x}\, {\rm tr} \/ [\epsilon_0^{(1)}
(\Pi_{(0)}^0 - \Pi_{(1)}^0) + \epsilon_0^{(2)} \Pi_{(1)}^0 +
\epsilon_i^{(1)} (\Pi_{(0)}^i - \frac12 \epsilon^{ij} A_j^{(0)} - 
\Pi_{(1)}^i - \frac12 \epsilon^{ij} A_j^{(1)}) +
\epsilon_i^{(2)} (\Pi_{(1)}^i - \frac12 \epsilon^{ij} A_j^{(1)})]$
where
$\epsilon_\mu^{(1)}$ and $\epsilon_\mu^{(2)}$ are local
parameters. Performing the St\"ukelberg method and identifying
$\epsilon_\mu^{(2)}$ with $A_\mu^{(2)}$ again, we obtain the action
corresponding to the second procedure of the BFT formalism. Note that we
need not identify $\epsilon_\mu^{(1)}$ with another vector field since
it cancels out under the transformation in this second step. After
all, the infinite number of St\"ukelberg substitution is expected to
yield the desired action.

\section*{Acknowledgments \hfill{ }}

We would like to thank Y.-W. Kim, C.-Y. Lee, and Y.-J. Park
for useful discussions.
We thank also to J.J. Oh for collaboration at the
first stage of this work. The present study was in part supported by
the Basic Science Research Institute Program, the Ministry of
Education, Project No. BSRI-98-2414, and in part supported by the
Korea Science and Engineering Foundation through the Center for
Theoretical Physics in Seoul National University.


\end{document}